\documentclass[10pt]{iopart}

\usepackage{epsfig,iopams}
\usepackage{graphicx}   
\usepackage{dcolumn}    
\usepackage{bm}         
\usepackage{color,multirow,amstext}
\usepackage{threeparttable}
\usepackage{hyperref}
\hypersetup{
  colorlinks=true,
  linkcolor=blue,
  ps2pdf
}

\newcommand{\eqref}[1]{(\ref{#1})}

\begin{document}
\title[Spatial heterogeneity of W transmutation in a fusion device]{
Spatial heterogeneity of tungsten transmutation in a fusion device}
\author{M.R. Gilbert, J.-Ch. Sublet, and S.L. Dudarev}
\address{CCFE, Culham Science Centre,
Abingdon, Oxfordshire OX14 3DB, UK.}
\ead{mark.gilbert@ukaea.uk}
\date{\today}
\begin{abstract}
Accurately quantifying the transmutation rate of tungsten (W) under neutron irradiation is a necessary requirement in the assessment of its performance as an armour material in a fusion power plant. The usual approach of calculating average responses, assuming large, homogenised material volumes, is insufficient to capture the full complexity of the transmutation picture in the context of a realistic fusion power plant design, particularly for rhenium (Re) production from W. Combined neutron transport and inventory simulations for representative {\it spatially heterogeneous} models of a fusion power plant show that the production rate of Re is strongly influenced by the local spatial environment. Localised variation in neutron moderation (slowing down) due to structural steel and coolant, particularly water, can dramatically increase Re production because of the huge cross sections of giant resolved resonances in the neutron-capture reaction of \(^{186}\)W at low neutron energies. Calculations using cross section data corrected for temperature (Doppler) effects suggest that temperature may have a relatively lesser influence on transmutation rates.

\end{abstract}


\maketitle
\ioptwocol
The change in composition, or transmutation, of a material under neutron irradiation can significantly alter its structural, mechanical, and even thermodynamic properties. Tungsten (W) is the main candidate material for components predicted to experience high heat and neutron fluxes in conceptual designs of fusion power plant reactors. In particular, W is expected to withstand neutron fluxes with peak-energies of 14~MeV and fluxes of the order of \(10^{15}\)~n~cm\(^{-2}\)~s\(^{-1}\)~\cite{gilbertsublet2011}, and thermal loads from plasma exhaust in the divertor that could reach 10~MW~m\(^{-2}\)\cite{maisonnieretal2007}.
Crucially, W must maintain good thermal conductivity to allow efficient cooling, and be radiation-hard by resisting, as far as is possible, irradiation-induced embrittlement that could lead to structural failures. Both of these properties can be detrimentally altered due to the accumulation of transmutation impurities under neutron irradiation \cite{hofmann2015}. It is therefore important to accurately predict, via modelling and simulation, the expected transmutation rates in W during reactor operation in the context of a realistic fusion power plant design.

\begin{figure}[t]
{\includegraphics[width=0.49\textwidth,clip=true,trim=0cm 0cm 0cm
0cm]{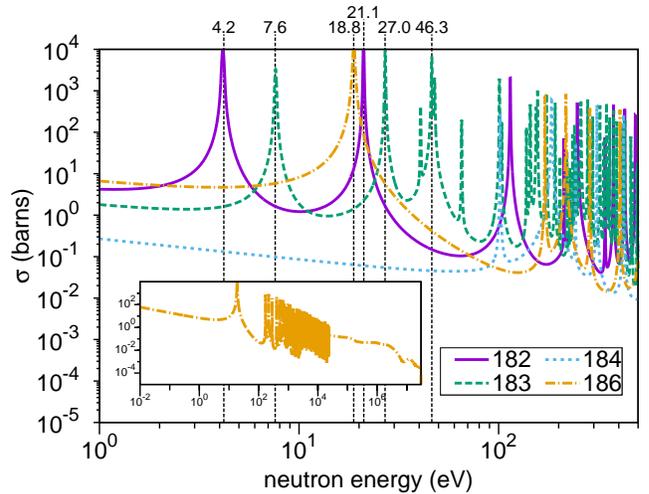}}
\caption{\label{xss}(Colour online) TENDL-2015~\cite{tendl2015} neutron cross-sections as a function of neutron energy for the neutron-capture (\(n,\gamma\)) reactions of the four main naturally occurring isotopes of W (referred to by their mass number in the key). The main figure focusses on the resolved resonance energy range, particularly that of the giant resonances below \(\sim\)30~eV. The inset shows only the \(^{186}\)W capture cross-section, but over a much wider energy range, including the unresolved resonance region between 22.5~keV and 121.4~keV, where the cross section appears as a smooth line. Above the resonance regions the four cross sections are virtually indistinguishable.}
\end{figure}

Previous studies~\cite{gilbertsublet2011} have investigated the transmutation rates in W under fusion conditions. That work considered the burn-up of W in a region of the fusion-plasma-exposed first wall of a power plant, and focussed on accounting for the so-called self-shielding effect, which is a particular issue for W. This phenomenon primarily concerns the giant (resolved) resonances of neutron-capture (\(n,\gamma\)) reactions. Figure~\ref{xss} shows the energy-dependent capture cross-sections \(\sigma \) (effectively ``reaction likelihood'') of the four main naturally occurring isotopes of W, where the giant resonances are clearly visible in the 1-30~eV neutron energy range for \(^{182}\)W (26.5~atm.\% of natural W), \(^{183}\)W (14.31\%), and \(^{186}\)W (28.43\%), but absent for \(^{184}\)W (30.64\%). A significant proportion of the fusion neutrons are absorbed in these resonances as they slow-down in the material, causing localised depletion of neutron energy fluxes (see, for example, in figure~\ref{spectra}), which in turn reduces the rate of other reactions with significant cross section in the same energy region -- hence ``self-shielding''. In the case of pure W the affected reactions are the capture reactions involving other isotopes of W and also isotopes of impurity elements (Re, Ta, Os, etc.) produced via transmutation. Other nuclear reactions generally have small cross sections at these lower neutron energies. The resulting influence on, in particular, the capture reaction rate (RR) of \(^{184}\)W and \(^{186}\)W is an important determining factor for the rate of transmutation to Re.

Rhenium is primarily created by the transmutation to, and subsequent \(\beta\)-decay of, the unstable \(^{185}\)W (half-life \(T_{1/2}=75\)~days) and \(^{187}\)W (\(T_{1/2}=24\)~hours) nuclides produced from both the (\(n,\gamma\)) reactions illustrated in Fig.~\ref{xss} and the (\(n,2n\)) reaction on \(^{186}\)W (see later).

Properly accounting for the resonances in neutron transport simulations requires consideration of two factors. Firstly, the simulated neutron fluxes must accurately account for the flux depletion at and below the giant resonances in the neutron energy spectra. Secondly, the simulation of the nuclide inventory evolution, which is most often accomplished using a numerical solver such as FISPACT-II~\cite{subletetalnds2017,fispact}, should correctly include the RR contributions from the giant resonances.

However, properly including these contributions in the total reaction rate RR\(^{tot}\) per atom, which is defined via
\begin{equation}
\texttt{RR}^{tot}\equiv\int_{0}^{\infty}\texttt{RR}(E)dE=\int_{0}^{\infty}\sigma(E)\phi(E)dE,
\end{equation}
for cross section \(\sigma(E)\) and flux \(\phi(E)\) functions of neutron energy \(E\), is complicated by the fact that the neutron fluxes, and hence the nuclear cross-section data, are typically represented not by integrals but by finite sums using an energy bin structure. This is a computational necessity when ``tallying'' neutron fluxes in a particle transport code such as MCNP6~\cite{mcnp1}. Then, for a bin structure containing \(N\) groups, the total reaction rate per atom is written as
  \begin{equation}\label{groupwiseeq}
\texttt{RR}^{tot}\approx\sum_{i}^{N}\sigma_i\phi_i,
\end{equation}
where \(\sigma_i\) and \(\phi_i\) are the total cross-section and flux, respectively, in energy bin \(i\). The problem arises because the cross sections of the sharp (in energy) resonances are overestimated by the coarsening procedure that produces nuclear reaction data in a particular energy-bin structure (see, for example, fig.2b in~\cite{gilbertsublet2011}, and figure~\ref{xss186} below). This can, in turn, produce an overestimation in RR\(^{tot}\) for the capture reactions. The effect is particularly dramatic in W because of the huge cross sections and their variation associated with the giant resonances, which, as figure~\ref{xss} shows, are many orders of magnitude higher than the cross sections at surrounding neutron energies.

Of course, it is theoretically possible to use an arbitrarily fine bin structure to try and properly represent the resonances. In modern computing, the total number of energy groups \(N\) used to represent the spectrum can be of the order 1000 or more, but even this is not enough, especially if the unresolved resonance region is to be properly represented as well. For example, in uranium (U), where self-shielding is also important, it was shown~\cite{Haeck,743104} that around 43000 bins would be required to accurately compute the RR.

The standard way to overcome the limitations of the energy bin approach for self-shielding is to apply self-shielding factors (SSFs) to correct the overestimation of RRs. For example, figure~\ref{xss186} exemplifies how such corrections would adjust \(^{186}\)W(\(n,\gamma\)) to either the infinitely dilute case (often applied in simulations), or to 50~barns, which is a typical value for the composition of pure W and at room temperature.
In FISPACT-II~\cite{subletetalnds2017,fispact} SSFs can be calculated for any particular neutron flux spectrum and material combination using ``probability tables'' to define the correct dilution of the resonance cross-sections into the appropriate energy bin structure.
Probability table data sets are supplied with FISPACT-II
for various nuclear library and temperature combinations in the energy range from 0.1~eV up to the end of the unresolved resonance energy range (121~keV for \(^{186}\)W\((n,\gamma)\) in TENDL-2015~\cite{tendl2015}) in the same fine UKAEA-709~\cite{subletetalnds2017} group structure as
the cross sections. Probability
table forms are also employed by Monte Carlo codes, such as TRIPOLI and MCNP6~\cite{MaBlCuSu09}, in the unresolved energy range, and by the fast deterministic
code ERANOS~\cite{Ruggieri}, but with some restrictions in both
resolved and unresolved resonance energy ranges.

\begin{figure}[t]
{\includegraphics[width=0.49\textwidth,clip=true,trim=0cm 0cm 0cm
0cm]{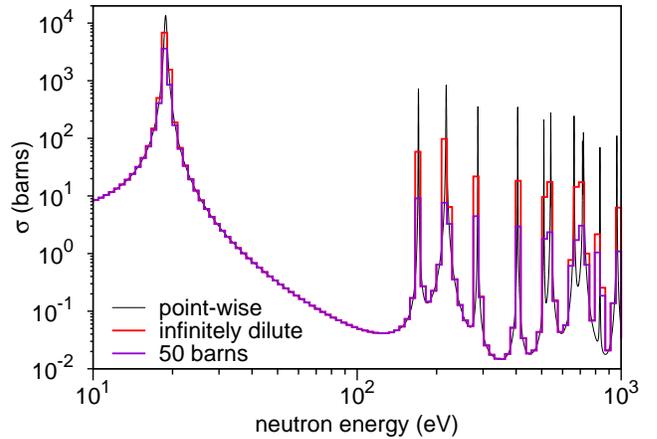}}
\caption{\label{xss186}(Colour online) Neutron-capture cross-section of \(^{186}\)W showing, for the resolved resonances, the impact of self-shielding factors (SSFs) on the conversion of the original point-wise data to group-wise format at two dilutions: infinitely dilute and at 50~barns (a typical value for the cases in this work -- see~\cite{MaBlCuSu09,fispact} for more details).}
\end{figure}

In~\cite{gilbertsublet2011}, where the goal was to compare the transmutation rate of W to those of other materials in the same neutron flux spectrum, this was generalised to find SSFs that not only corrected the overestimation of the resonance peaks, but also accounted for the flux depletions, which would otherwise only be present if the fluxes were obtained from a computationally demanding neutron transport simulation performed for W in the correct geometry. Depth-averaged SSFs (the average of SSFs computed as a function of depth into 30~cm of a tungsten-water mixture) were evaluated for the important neutron-capture reactions. In this case the SSFs were the ratio of the RR\(^{tot}\) value obtained from Monte Carlo simulations using continuous point-wise cross-section data, to the overestimated RR\(^{tot}\) obtained in FISPACT-II using group-wise nuclear data. The approach resulted in SSFs that reduced the RR\(^{tot}\) for (\(n,\gamma\)) on \(^{186}\)W by more than 90\%.



However, the work in~\cite{gilbertsublet2011} and in other studies of transmutation in W~\cite{subletsawan1999,cottrelletal2006,sawanfst2014} consider {\it volume-averaged} neutron fluxes for a particular reactor region. Specifically, all of the studies~\cite{gilbertsublet2011,subletsawan1999,cottrelletal2006,sawanfst2014} used a neutron spectrum that was averaged over the full depth of an homogenised first wall armour tile, which is typically of the order of 1-2~cm thick~\cite{gilbertetalJNM2013}. While this approach produced the correct homogenised, average transmutation result for the armour component, the thicknesses considered are much greater than those typically used in experimental tests, or in investigations of microstructural changes induced by irradiation. Furthermore, variations on a much finer scale are likely to significantly influence certain properties, such as thermal conductivity and sputtering yields. It is therefore necessary to consider the variation in the transmutation of W on much finer length scales than has been investigated previously. The study described below shows how the local environment around W can significantly alter the amount of Re produced under fusion neutron irradiation, to an even greater degree than the aforementioned SSFs.

\begin{figure*}[t]\centering
{\includegraphics[width=0.25\textwidth,clip=true]{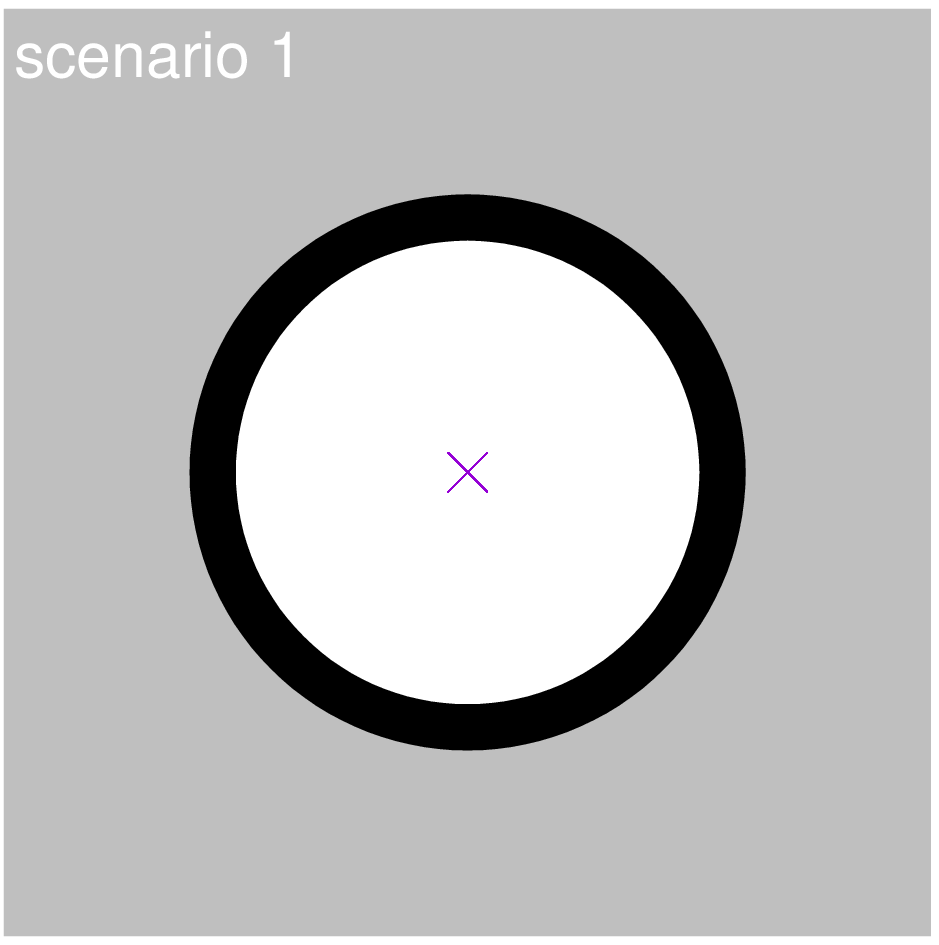}}
{\includegraphics[width=0.25\textwidth,clip=true]{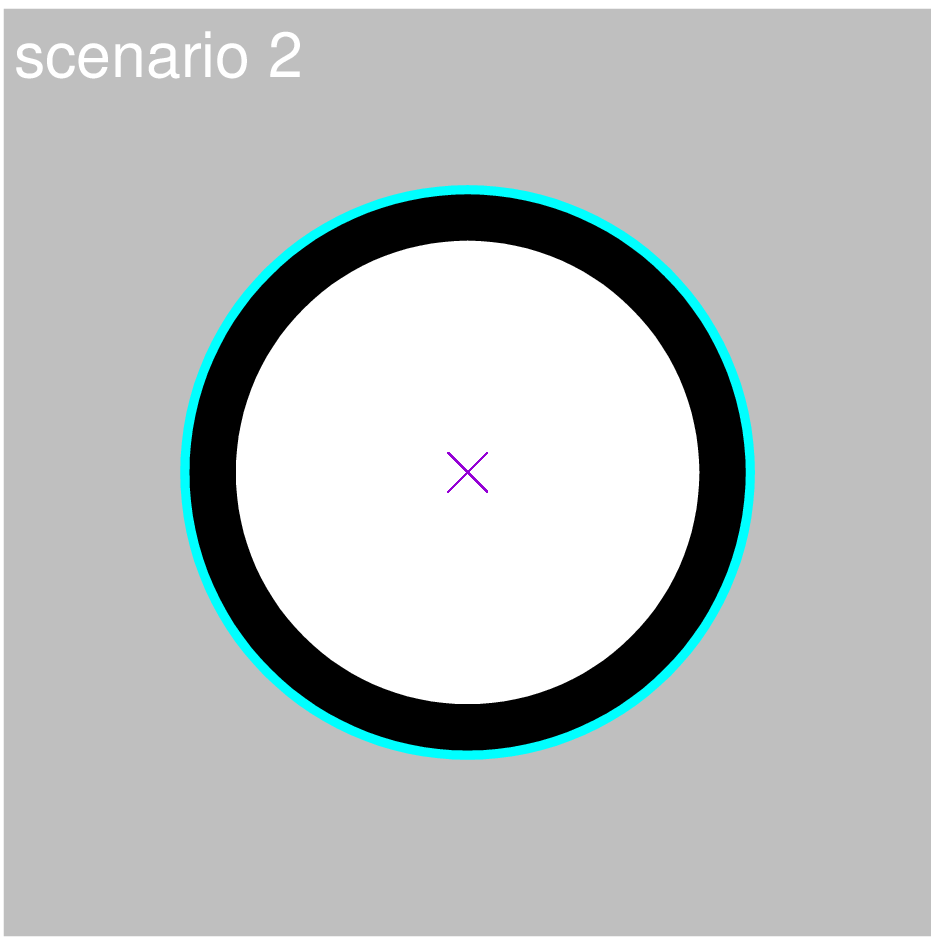}}
{\includegraphics[width=0.25\textwidth,clip=true]{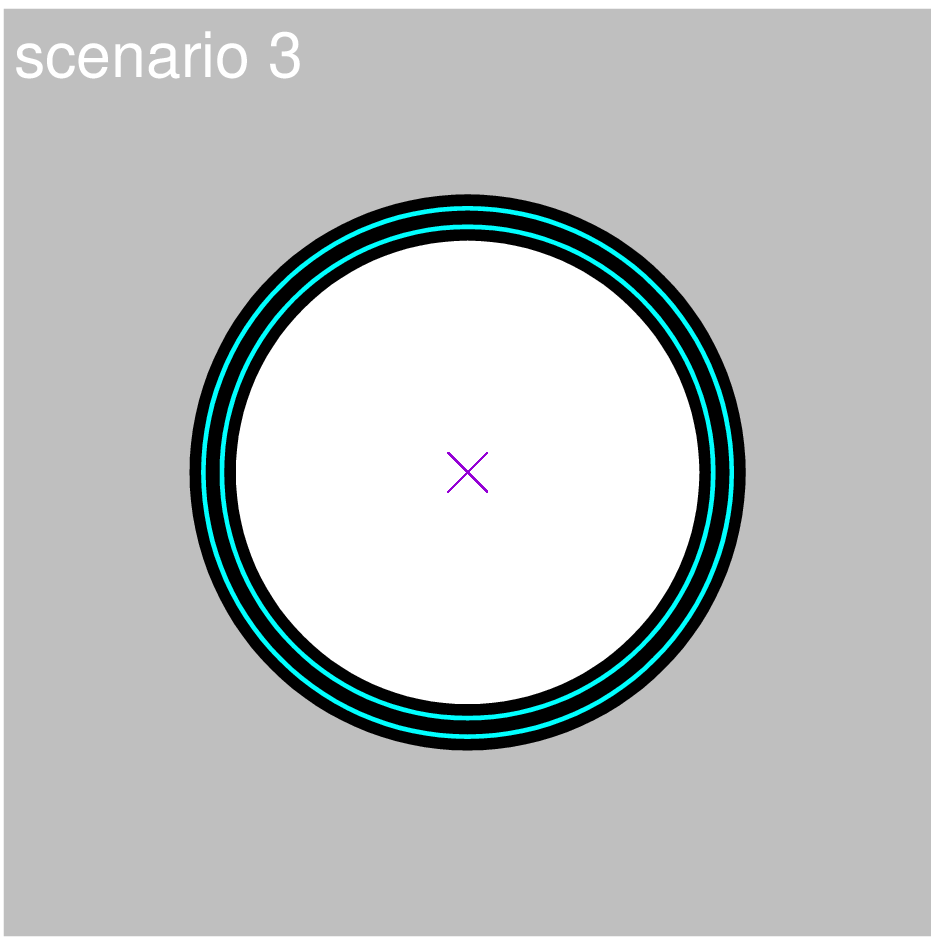}}
\caption{\label{geometry}(Colour online) Simulation set-ups for the three model scenarios. Black: pure tungsten; Grey: 80-20 by volume steel-water mixture; Blue: water. White: Vacuum.}
\end{figure*}

\begin{figure}[h]
{\includegraphics[width=0.49\textwidth,clip=true,trim=0cm 0cm 0cm
0cm]{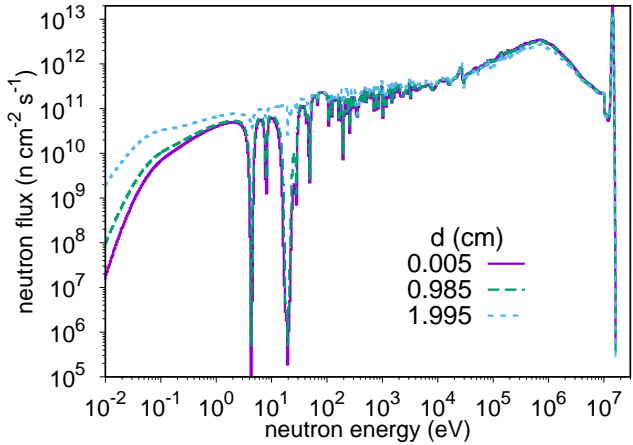}}
\caption{\label{spectra}(Colour online) Flux spectra simulated in three of the 0.1~mm W layers in the scenario~1 geometry (see figure~\ref{geometry}). The legend label for each spectrum is the (midpoint) depth (d) into the W of the corresponding layer. Note that the statistical errors from the MCNP6 simulations are not included here as they are too small to be distinguished from the lines.}
\end{figure}

To investigate and exemplify the importance of performing transmutation simulations in a fully spatially resolved geometry, in this letter we consider a simplified model to represent a typical fusion environment for a tungsten armour tile.
The model consisted of a 2~cm thick spherical shell of pure W (density 19.3~g~cm\(^{-3}\)) with an inner radius of 10~cm, surrounding a vacuum. The ``scenario 1'' graphic in figure~\ref{geometry} depicts the model set-up. A SS316 steel and water 80\%-20\% by volume mix surrounded the steel out to a radius of more than 30~cm, to represent the typical moderating material that would be present behind a tungsten armour tile in a fusion reactor. Any neutrons that exited this extended moderator region were terminated (i.e. not reflected back). At the centre of the geometry an isotropic point source was used, with an energy distribution corresponding to a 14~MeV neutron source for a deuterium-tritium (DT) fusion plasma at \(T=20\)~keV.

The tungsten shell was split into a fine-resolution spatial grid in 0.1 mm intervals, and a volume-averaged neutron flux tally was recorded in each by MCNP6~\cite{mcnp1} in an  energy-bin structure containing 660 bins below 30~MeV~\cite{fispact}. Even in such a simplified geometry, simulating the propagation of neutrons through tungsten is relatively computationally intensive. A suitable particle-weight biasing mesh was calculated using ADVANTG~\cite{advantg} to reduce computational effort required in MCNP6 to produce reasonable statistics. \(10^{10}\) neutron histories were sampled for the scenario 1 geometry, which took around 6 weeks on 32 cpus, resulting in statistical uncertainties of less than 1\% in the majority of energies bins above 0.1~eV (less than 2\% above 1e-3~eV). The few exceptions to this were in the spectra of the W layers closest to the neutron source, where bins containing the giant resonances suffered from insufficient sampling, but those containing the resonance of the important (for transmutation to Re) \(^{186}\)W capture reaction still had uncertainties of less than 10\% in these cases.

Note that MCNP6 uses continuous-energy cross-section data and so can accurately model resolved resonances without the need for any special computational treatment. However, in the unresolved resonance range of the capture reactions on W isotopes, the continuous cross-sections appear as smooth functions of energy because the resonances are too close together. MCNP6 properly accounts for the resonance self-shielding in this unresolved range using a similar probability table approach as used by FISPACT-II, but instead applies a statistical sampling approach (see~\cite{MaBlCuSu09,mcnp1} for more details).

Figure~\ref{spectra} shows three of the spectra produced from these simulations: the first, the last, and a middle 0.1~mm layer. The ``per source neutron'' tally results from MCNP6 have been normalised according to the source rate (n~s\(^{-1}\)) required to produce 2~MW~m\(^{-2}\) of 14~MeV neutron wall loading on the internal face of the W shell, which is a typical fusion reactor value~\cite{maisonnieretal2007}. The self-shielding ``troughs'' are clearly apparent in each spectrum, even in the plasma-facing layer, which nonetheless experiences a significant flux of moderated neutrons that have backscattered from deeper regions. Note that in the last layer, closest to the steel-water moderator, the backscattered neutrons created by moderation and multiplication provide additional neutrons into these depleted energy regions and thereby reduce the troughs.

The observation of self-shielding depletions at all depths in the W demonstrates that this might not be a particularly significant factor as far as variations in transmutation rate is concerned. It appears that there is sufficient backscatter of neutrons to populate the energy ranges of the resonances at all depths.

Although the total neutron flux drops slightly with depth into the W -- from \(3.2\times10^{14}\)~n~cm\(^{-2}\)~s\(^{-1}\) in the first layer to \(2.5\times10^{14}\)~n~cm\(^{-2}\)~s\(^{-1}\) at the back -- figure~\ref{spectra} also shows that the flux of lower energy neutrons, primarily responsible for the W-Re transmutations, is highest in regions closest to the moderator due to backscattering.

Each of the calculated spectra have been used in separate inventory simulations with FISPACT-II, using the latest TENDL-2015~\cite{tendl2015} nuclear cross-section libraries. Pure W was irradiated for 2~fpy (full power years) irradiation, which is representative of the expected exposure lifetimes of first wall components in fusion power plants. The probability-table approach was used in each FISPACT-II simulation to model the dilution effects associated with the resonances (both giant-resolved and unresolved) in the capture cross-sections of each W isotope in the starting mixture (180, 182, 183, 184, and 186), and hence to compute SSFs to scale the RR\(^{tot}\) values computed using the group-wise flux and cross section data (equation~\eqref{groupwiseeq}). For the \(^{186}\)W(\(n,\gamma\))\(^{187}\)W reaction, which accounts for 100\% of the stable \(^{187}\)Re isotope produced in the simulations (see later) the effective SSFs applied to RR\(^{tot}\) ranged from 0.56 to 0.68 (56 to 68\% reduction). Similarly, for the \(^{184}\)W(\(n,\gamma\))\(^{185}\)W reaction, which also contributes to Re production, the effective SSFs ranged between 0.43  and 0.58.

\begin{figure}[h]
{\centering{\includegraphics[width=0.49\textwidth,clip=true,trim=0cm 0cm 0cm
-2cm]{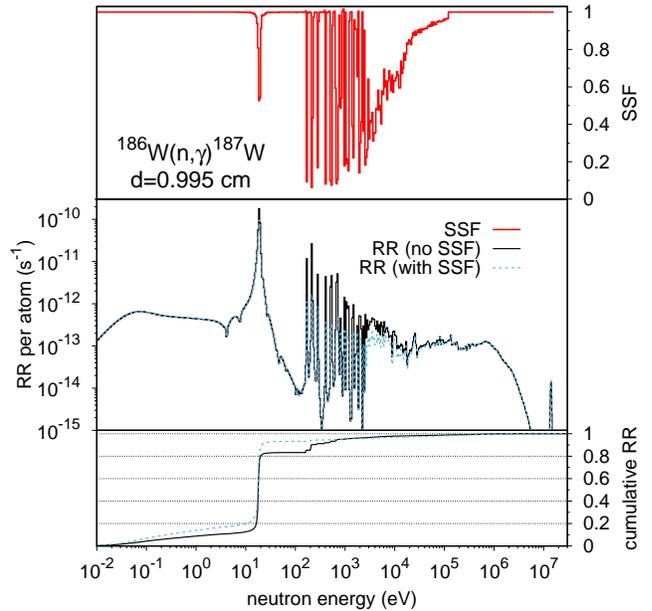}}}
\caption{\label{RR186}(Colour online) Corrected and uncorrected RR values, corresponding SSFs, and cumulative RRs as a function of neutron energy for the flux spectrum in the final 0.1~mm layer of the W in scenario 1 and the neutron-capture reaction on \(^{186}\)W.}
\end{figure}

The \emph{effective} SSFs above were computed (in FISPACT-II) as the ratio of the new RR\(^{tot}\) to the old, uncorrected value. For a particular reaction the corrected RR\(^{tot}\) is obtained as the sum of individual corrected RR values in each neutron energy bin (other approaches are available in FISPACT-II -- see~\cite{fispact}). It is instructive to examine the variation of the corresponding SSF values as a function of energy bin to observe where the main adjustments are made. Figure~\ref{RR186} plots the SSF variation, and the corrected and uncorrected RRs as a function of energy, for the neutron-capture reaction on \(^{186}\)W in the neutron flux spectrum calculated for the final 0.1~mm layer of the W (see figure~\ref{spectra}).


Figure~\ref{RR186} shows that the largest corrections to the RR are actually associated with the unresolved resonances (cf. figure~\ref{xss} inset). However, the corrected RR associated with the giant resonance at 18.8~eV~\cite{gilbertsublet2011} dominates the reaction rate in this layer of the W -- as shown by the cumulative RR plots at the bottom of the figure, where around 70\% of the RR\(^{tot}\) originates from the giant resonance. On the other hand, for example, in the first 0.1~mm layer, the corrected contribution from the giant resonance is almost negligible (see figure~\ref{RRs_comp}), which is a result of the combined influence of the self-shielding flux depletions and the (relatively) hard, unmoderated neutron spectrum at this depth (discussed further in the reaction path analysis below).

\begin{figure}[h]\centering
{\includegraphics[width=0.49\textwidth,clip=true]{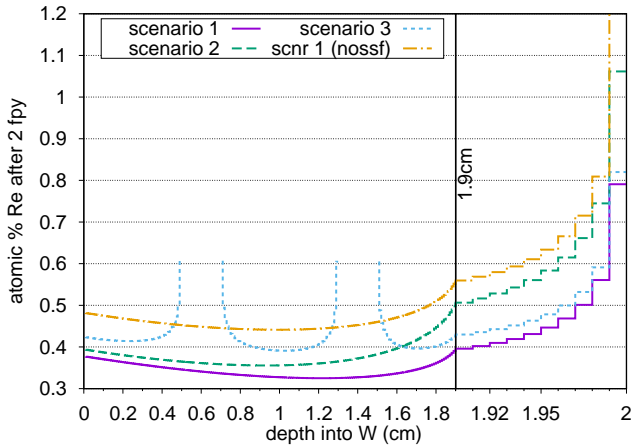}}
\caption{\label{re_rt}(Colour online) The variation with depth into W of transmutant Re concentration after simulated 2~fpy irradiations for the three different model scenarios considered. Inset figure: Ta concentration with depth profiles from the same inventory simulations. The results are plotted as step functions with a step for each 0.1~mm W slice. The final 1~mm is plotted on a finer length to make the large variation in these depths easier to appreciate. Also shown is the scenario~1 result without SSFs.}
\end{figure}

Figure~\ref{re_rt} shows the final Re concentrations from the inventory simulations as a function of depth into the W after the 2~fpy irradiations. Also shown is the scenario-1 result that would have been obtained without the SSFs discussed above, i.e. without corrections for the dilution of giant resonances in an energy bin structure. What is most striking about the comparison between the two scenario-1 curves is the fact that the difference between them is completely swamped by the variation in Re concentration observed at the back of the W, next to the steel-water moderator. While it is true that the Re concentration would have been over-predicted without the corrections, by between 25 and 50\%, it is also the case that the variations caused by the local spatial environment of the model are much greater. The difference between the maximum (at the back) and minimum (at a depth of around 1.2~cm) Re concentrations after 2~fpy is more than 4600 atomic parts per million (appm) or 0.46~atm.\%, which could result in a profound difference in how the material behaviour is altered \cite{hofmann2015}. The volume-averaged Re concentration, obtained using a neutron flux spectrum calculated for the entire depth of W, is around 3500~appm, while the standard deviation of the spatially-resolved values from this ``average'' is more than 400~appm.

Using the reaction pathway analysis features of FISPACT-II (see~\cite{subletetalnds2017}) for details) it is possible to gain further insight into the reasons for the variation seen in figure~\ref{re_rt}. The RR of three key reactions determines the amount of Re produced from W under neutron irradiation. Two of these are the aforementioned neutron-capture (\(n,\gamma\)) reactions on \(^{186}\)W and \(^{184}\)W. The other is the neutron multiplication reaction (\(n,2n\)) on \(^{186}\)W, which leads to \(^{185}\)Re via decay of \(^{185}\)W (the same route as follows \(^{184}\)W(\(n,\gamma\))).

\begin{figure}[h]\centering
{\includegraphics[width=0.49\textwidth,clip=false]{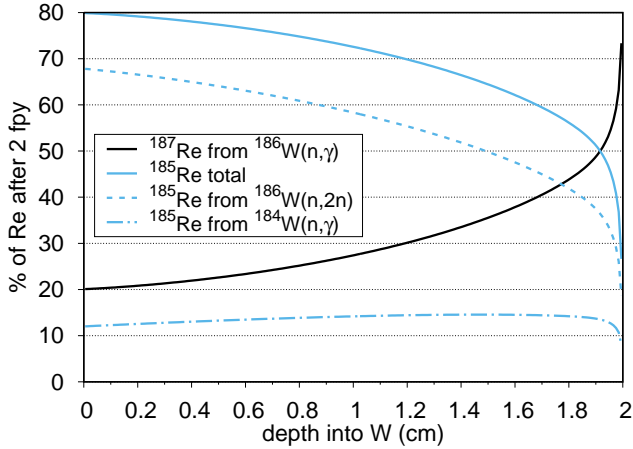}}
\caption{\label{re_paths}(Colour online) Fractional contribution depth-profiles for the primary \(^{187}\)Re and \(^{185}\)Re isotopes in transmutant Re from W in the scenario 1 model (solid curves). For \(^{185}\)Re, the contribution is also separated into its two main production channels.}
\end{figure}

\begin{figure}[h]
{\centering{\includegraphics[width=0.49\textwidth,clip=true,trim=0cm 0cm 0cm
0cm]{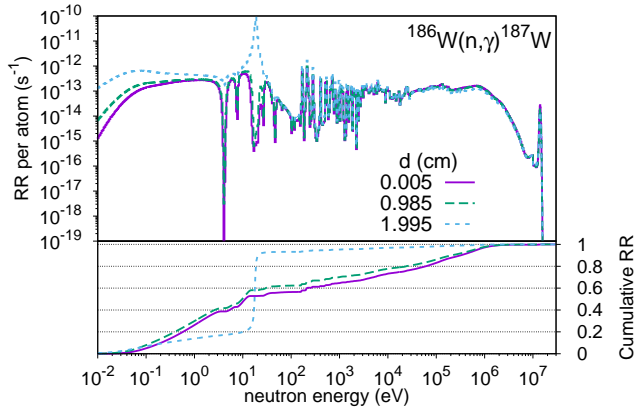}}}
\caption{\label{RRs_comp}(Colour online) SSF-corrected RRs and cumulative RRs as a function of neutron energy for \(^{186}\)W\((n,\gamma\)) at three different 0.1~mm W layers in the scenario 1 geometry. The legend label for each spectrum is the (midpoint) depth (d) into the W of the corresponding layer.}
\end{figure}

Figure~\ref{re_paths} plots (as solid lines) the fractional contributions from \(^{185}\)Re and \(^{187}\)Re to the total Re content predicted by the inventory simulations. The total fractional contribution from these two curves is virtually 100\% at all depth into W, confirming that these are the only Re isotopes produced in significant concentrations under neutron irradiation of W. However, the relative proportion of these two nuclides changes dramatically with depth.

At shallow depths \(^{185}\)Re comprises more than 75\% of the total Re created, and furthermore, most of that is produced via the \((n,2n\)) reaction on \(^{186}\)W, whose specific contribution is also shown in the plot (dashed line). However, as the depth into W increases, two changes are apparent. Firstly, the proportion of  \(^{185}\)Re that comes from \(^{186}\)W\((n,2n\)) drops; and secondly, the proportion of Re that is \(^{187}\)Re increases. The \(^{186}\)W\((n,2n)\) change is understandable given that the reaction has a threshold at 7.2~MeV. As the neutron spectrum becomes more moderated with depth, which, in particular, means that there is a reduction in the neutrons in the MeV energy range, this reaction becomes less likely -- RR\(^{tot}\) for this reaction falls by almost 50\% between the front and back layers.

At the same time, however, and as has already been noted, the total neutron flux does not fall very much with depth, and so in the neutron spectra of deeper W layers most of the high-energy neutrons have been replaced by lower energy ones, which is often referred to as a ``spectral shift''. This causes the RR\(^{tot}\) of \(^{186}\)W\((n,\gamma\)), which is responsible for all of \(^{187}\)Re,  to increase by a factor of 7 between the front and back, because of its giant resonance. Note, on the other hand, for \(^{184}\)W\((n,\gamma\)), which has no giant resonances, RR\(^{tot}\) is virtually constant with depth (dash-dot curve in figure~\ref{re_paths}).

Figure~\ref{RRs_comp} plots the \(^{186 }\)W\((n,\gamma\)) RR and cumulative RR as a function of energy for three different 0.1~mm layers of the W (the same first, last, and middle considered in figure~\ref{spectra}), demonstrating how the giant resonance at 18.8~eV dominates in the layers close to the moderator, while it has  negligible contributions at shallower depths.


The results above have highlighted the importance of considering the proper geometry when predicting transmutation, particularly for transmutants whose production routes involve reactions that are very sensitive to lower energy neutrons. Results for two other model geometries, scenarios 2 and 3, illustrate this still further. Scenario 2 considers a situation where there is a cooling channel of water directly behind the tungsten, modelled here as a 4~mm shell between the W and the steel-water mixture -- see figure~\ref{geometry}. In the neutron transport simulations this has the effect of increasing neutron (back) scattering and moderation, producing increased fluxes of low energy (sub keV) neutrons through a significant proportion of the W, and leading to an increase in the Re production compared to scenario 1 by more than 30\% at the back of the W, and around 15\% 50~mm away from the water (see figure~\ref{re_rt}). Near the front (relative to the neutron source) of the W, however, there is hardly any change in Re production.

Meanwhile, Scenario 3 instead considers a model where water channels are embedded in the W, which may be unavoidable in a real reactor due to the severe heat loads expected (particularly in divertor regions). In this case two 4~mm layers of water were inserted into the original scenario 1 model, at equal distances from the edges of the W and each other (see figure~\ref{geometry}). As with scenario 2, the water produces additional neutron moderation and scattering into the adjacent W layers, thereby increasing the RR of the neutron-capture reactions (the threshold (\(n,2n)\) reaction meanwhile is largely unaffected). In figure~\ref{re_rt}, the resulting localised increases in Re concentrations clearly mark the locations and boundaries of the water channels.

Another striking feature of the modelling results is how little the different scenarios influence the concentration of Ta, which is another of the main transmutation product in the simulations. Os is also produced at concentrations of the order of a few hundred appm, but the detailed variation in its production, which would require SSF corrections for (initally) unknown concentrations of Re isotopes, is beyond the present scope.

\begin{figure}[h]\centering
{\includegraphics[width=0.49\textwidth,clip=true]{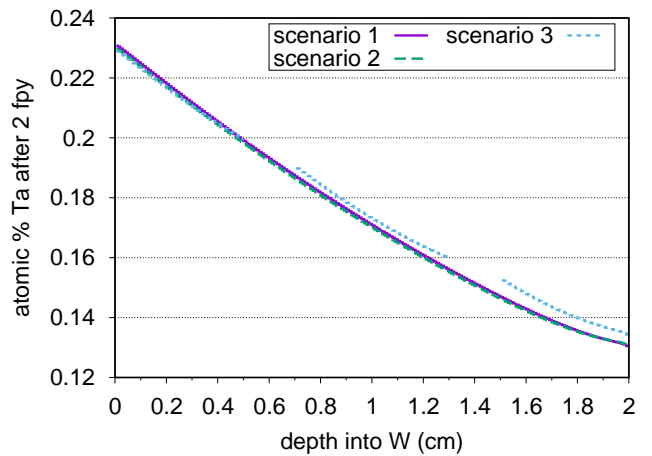}}
\caption{\label{ta_rt}(Colour online) The variation with depth into W of transmutant Ta concentration after simulated 2~fpy irradiations for the three different model scenarios considered.}
\end{figure}

Figure~\ref{ta_rt} shows the depth profile of Ta concentrations in the three scenarios (with SSFs, although this has no significant impact on Ta production). The three curves are almost indistinguishable and follow a downward trajectory. Again, this is easily understood by realising that more than 90\% of the Ta at all depths is produced via \(^{182}\)W(\(n,2n\))\(^{181}\)W(\(\beta^+\))\(^{181}\)Ta, where the (\(n,2n\)) reaction in this case has an even higher threshold than \(^{186}\)W(\(n,2n\)) at 8.1~MeV. As the fluxes of high energy (above 1~MeV) neutrons decreases through moderation, so does the RR\(^{tot}\) of such \((n,2n)\) reactions, and none of the moderation and scattering at lower neutron energies in the water or steel has any impact on this. This is why, in figure~\ref{ta_rt}, even the interior water channels of scenario~3 have barely any impact on the concentration profile.

FISPACT-II can also quantify the variation in damage dose in the inventory simulations for scenario 1, using the standard displacements per atom (dpa) measure. The dpa-profile with depth is similar to that for Ta concentration -- meaning that only the variation in fluxes at MeV neutron-energies has any influence. This trend is also seen in radioactivity at medium and long decay times following the 2~fpy irradiation, where isotopes of Ta are the main contributors to the total activity (measured in Bq~kg\(^{-1}\)) in the W. On the other hand, at short timescales, immediately after the 2~fpy irradiation, the activity is dominated by \(^{187}\)W (\(T_{1/2}=24\)~hours) and \(^{185}\)W (\(T_{1/2}=75\)~days), and so the variation in the neutron fluxes in the resonance range is again important. As already discussed for Re production, the variations in RR\(^{tot}\) for \(^{186}\)W(\(n,2n\)) and \(^{186}\)W(\(n,\gamma\)) determine the relative proportions of \(^{187}\)W and \(^{185}\)W, and hence short term activity, although in this case the levels (of activity) at the front and back of the W shell are more similar than was the case for the Re concentrations.


The importance of considering heterogeneity in the local environment around W for transmutation to Re  was recently confirmed by direct experimental observations~\cite{klimenkovetal2016,abernethyarmstrong2017}. A sample of W was irradiated in the high-flux reactor (HFR) at NRG, Petten in the Netherlands for 208 effective full-power days. A FISPACT-II simulation of this irradiation with a standard neutron flux spectrum for HFR resulted in a 4~atm.\% Re concentration at the end of the simulation (5\% without a self-shielding correction). However, an experimental analysis of the sample, using energy-dispersive X-ray spectroscopy (EDX) revealed that the surface Re concentration was in the range 1.2-1.4 atm.\%. Therefore, a more representative neutron transport simulation of the experimental set-up was performed, using the actual reactor environment in which the W samples were exposed to neutrons. The resulting neutron spectrum was somewhat different to the expected spectrum in the corresponding part of the reactor -- the flux in the thermal neutron region below 0.1~eV, in particular, was significantly lower. This led to RR\(^{tot}\) for the important \(^{186}\)W(\(n,\gamma\))\(^{187}\)W reaction being reduced by around 70\%. The subsequent inventory calculation this time predicted 1.4~atm.\% after the 208 days -- a remarkably good agreement with experimental measurements.


Another potential source of variation for nuclear reaction rates concerns the influence of temperature. All of the results discussed so far were obtained using reaction cross-sections calculated for room temperature (294~K), in both MCNP6 and FISPACT-II. However, a real reactor will operate at much higher temperatures, which can be factored-in when computing the cross sections, primarily via a doppler-broadening of the resonances.

\begin{figure}[h]\centering
{\includegraphics[width=0.49\textwidth,clip=true]{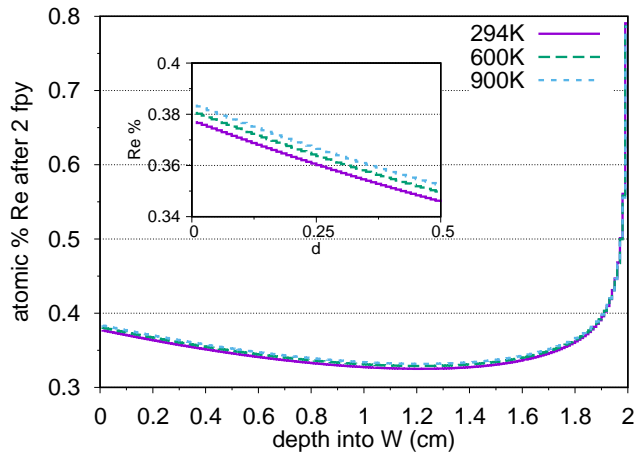}}
\caption{\label{re_temp}(Colour online) The variation with
depth into W of transmutant Re concentration after simulated
2 fpy irradiations for the scenario 1 model (see figure~\ref{geometry}) at three different nuclear cross-section temperatures. The inset figure is a zoom of the curves over the first 0.5~cm of the W.}
\end{figure}

Both the neutron transport and inventory calculations for scenario 1 have been repeated using nuclear reaction data at two alternative temperatures -- 600 and 900~K. Figure~\ref{re_temp} shows the Re concentration-with-depth profile results for these two new cases, as well as the original results corresponding to room temperature. In this case, the variation with temperature is only relatively small, particularly in comparison to the variation with depth at a given temperature. The (slight) trend with increasing temperature is the expected increase in transmutation to Re caused by the broadening of the giant resonances and hence an increase in the neutron-capture RRs.

In summary, in this paper we have shown how the standard practice of computing compositional changes due to nuclear transmutation using neutron flux spectra averaged over too large a volume can lead to misleading predictions for W. Fine length-scale simulations in a simplified geometry demonstrate that the local neutron environment can produce significantly spatially varying transmutation rates of W into Re -- over much shorter distances than those typically considered in homogenised neutron transport geometries for fusion. The geometry must be modelled at an appropriate spatial resolution to provide the full transmutation picture in W and to inform experimental testing, theoretical materials models, and engineering design decisions. Proper computational treatment of the giant resonances in neutron-capture cross-sections, via self-shielding correction factors, also influences the transmutation rates, although not as significantly as the spatial heterogeneity. The self-shielding phenomenon itself, where the resonances cause neutron flux depletion, appears to be relatively consistent at all depths in W because the resonance energy ranges are (re-)populated by neutron backscattering. Variation in Re production in W with temperature is relatively minor in our case.

\ack
The authors gratefully acknowledge the efforts of S. van der Marck to properly define the experimental history of the W samples irradiated in the HFR reactor at NRG, Petten.
This work was funded by the RCUK Energy Programme [grant number EP/I501045]. To obtain further information on the data and models underlying this paper please contact PublicationsManager@ukaea.uk

\section*{References}
\bibliographystyle{nuclear_fusion}
\bibliography{W_transmutation_revisited}
\end{document}